\documentclass[aps,showpacs,amsmath,pre]{revtex4}
\usepackage{graphicx}
\usepackage{float}
\newcommand{\y }{\'{\i}}
\newcommand{\be }{\begin{equation}}
\newcommand{\ee }{\end{equation}}

\begin{document}


\title{A model for the catalytic oxidation of CO, including gas-phase impurities and CO desorption}
\author{G.~M.~Buend\y a}
\affiliation{Physics Department, Universidad Sim\'on Bol\y var,\\
Apartado 89000, Caracas 1080, Venezuela}
\author{P.~A.~Rikvold}
\affiliation{ Department of Physics, \\
Florida State University, Tallahassee, FL 32306-4350, USA}

\date{\today}

\begin{abstract}
We present results of kinetic Monte Carlo simulations of a modified Ziff-Gulari-Barshad model for the reaction CO+O $\rightarrow$ CO$_2$ on a catalytic surface. Our model includes impurities in the gas phase, CO desorption, and a modification known to eliminate the unphysical O poisoned phase. The impurities can adsorb and desorb on the surface, 
but otherwise remain inert. In a previous work that did not include CO desorption [G.~M.~Buend\y a and P.~A.~Rikvold, Phys.~Rev.~E, {\bf 85} 031143 (2012)], we found that the impurities have very distinctive effects on the phase diagram 
and greatly diminish the reactivity of the system. If the impurities do not desorb, once the system reaches a stationary state, the CO$_2$ production disappears. When the impurities are allowed to desorb, there are regions where the CO$_2$ reaction window reappears, although  greatly reduced.
Following experimental evidence that indicates that temperature effects are crucial in many catalytic processes, here we further analyze these effects by including a CO desorption rate. We find that the CO desorption has the effect to smooth the transition between the reactive and the CO rich phase, and most importantly it can counteract the negative effects of the presence of impurities by widening the reactive window such that now the system remains catalytically active in the whole range of CO pressures. 
\end{abstract}


\maketitle

\section{Introduction}
\label{sec:I}

Heterogeneous catalysis presents a rich variety of non-equilibrium phenomena, such as kinetic oscillations, 
chaos, spatiotemporal pattern formation, and hysteresis phenomena \cite{baxt02,chri94, yama95, imbi95, marro99}.
Catalytic processes are ubiquitous in nature, and they also have wide technological and industrial applications.
The demand for new, inexpensive, and efficient catalytic systems requires a solid understanding of how industrial conditions 
affect these processes.  Experiments realized in ultrahigh vacuum greatly differ from their industrial counterparts, in which contaminants are an unavoidable part of the process. Typical catalytic poisons are sulfur compounds, hydrocyanic acid, mercury, compounds of phosphorous, etc.\ \cite{herz81, liu93, asak97}. The most common effect of such impurities is to decrease the number of reactive sites on the catalytic surface, thereby greatly reducing the efficiency of the process. Adsorption of impurities can be reversible or irreversible, and sometimes this distinction depends on the temperature \cite{forz99, yu98}. Studies of automotive catalysts indicate that sulfur poisoning is irreversible at temperatures below 
$650^{\circ}$C, but the catalytic activity can be restored at higher temperatures \cite{beck95}.
The present paper represents an extension of our previous studies \cite{buen12} of the effects of reversibly adsorbed 
impurities in the feed gas \cite{EIGE78,VIGI96} on a statistical mechanics model of 
the oxidation reaction of carbon monoxide (CO) on transition metal surfaces. 

This oxidation reaction is a prototypical 
surface reaction and has been extensively studied. Ultrahigh vacuum studies have established that Langmuir-Hinshelwood (LH, see Sec.~II) is the dominant mechanism for this reaction \cite{engl78}. In a seminal work, Ziff, Gulari, and Barshad 
(ZGB) proposed a minimalistic kinetic lattice-gas model that simulates the surface as a two-dimensional square lattice where the catalytic reaction occurs via a LH process \cite{ziff86}. In the ZGB model, CO molecules in the gas phase adsorb on single sites of the surface, while oxygen (O$_2$) molecules are dissociatively 
adsorbed on nearest-neighbor sites. Adsorbed nearest-neighbor CO and O react, producing carbon dioxide (CO$_2$) that instantly desorbs and leaves behind two empty sites. The process is controlled by a single parameter, $y$, that represents the probability that the next molecule arriving at the surface is CO. The parameter $y$ is proportional to the partial pressure of CO in the feed gas. This model exhibits two kinetic phase transitions: a continuous one at $y=y_1$, between an oxygen-poisoned state and a reactive state, and a discontinuous one at $y=y_2 > y_1$ between a reactive state and a CO poisoned phase. For $y_1<y<y_2$, the CO$_2$ production increases smoothly from zero at $y\approx y_1$, to its maximum value at $y \approx y_2$. For $y > y_2$, the reaction rate vanishes. 

For completeness we note that the ZGB model and its various extensions are not the only 
statistical mechanics based models for heterogeneously catalyzed CO oxidation. More detailed results are provided by 
several lattice-gas models that explicitly include energetic effects, such as adsorbate-adsorbate interactions and 
reaction and adsorption-desorption barriers obtained by quantum mechanical 
DFT calculations and/or comparison with experiments \cite{VOLK01,PETR05,LIU06,NAGA07,ROGA08,LIU09,HESS12}.

However simplified the ZGB model and its generalizations are, they provide a useful stage for testing the effects of 
various processes, alone and in combinations, on the reaction kinetics. 
In this paper we extend our previous work on a ZGB-like model that 
includes a reversibly adsorbed, inert species in the gas phase \cite{buen12}. The aspect that we add to our previous 
model, is the desorption of CO molecules \cite{kauk89,bros92,alba92}. This can be thought of as an effect of 
increased temperature. 

The original ZGB model has two particularly unphysical aspects, and several modifications have been suggested to 
remove these \cite{meak90, bros92, alba92, buen09}.
First, the oxygen-poisoned phase has not been observed in real systems because oxygen does not impede the adsorption of CO \cite{ehsa89, kris92, ertl90}. 
Here, we therefore remove this phase by modifying the adsorption mechanism of the oxygen molecule, such that the two adsorbed atoms enter two next-nearest-neighbor sites instead of two nearest-neighbor sites \cite{ojed12, buen12}. 
Physically, such an effect could be due to exclusion or repulsion between nearest-neighbor adsorbed O atoms 
\cite{BRUND84,JAME99,LIU99}, 
or a tendency of the O atoms to move apart once they reach the surface due to their thermal energy, known as 
hot-atom adsorption \cite{wint96, alba94, khan04}. 
This modification was already introduced in our previous study, Ref.~\onlinecite{buen12}.

Second, transitions from the phase of high to the phase of 
low CO coverage have been observed experimentally upon reduction of $y$ to below $y_2$ \cite{mats79}. 
Thus, the high-CO-coverage phase is not irreversibly poisoned. Reversibility of the discontinuous transition between the productive and the CO poisoned phase can be reproduced by including a positive CO desorption 
rate, mimicking the effect of nonzero temperature \cite{bros92,alba92, kauk89}. 
(Experimental evidence indicates that the desorption rate of oxygen is negligible \cite{ehsa89}.) 
Hysteresis is then observed as $y$ varies close to $y_2$, which now becomes a function of the desorption rate \cite{mach05a, tome93, buen06}. 
In the present paper we study the effects of this CO desorption process on our previous model for CO oxidation 
with inert impurities in the gas phase and next-nearest-neighbor adsorption of oxygen atoms \cite{buen12}. In the next paragraph we give a brief description of this model. 

Recent experiments on CO oxidation on Au nanoparticles in the presence of  impurities show that the system becomes catalytically active at high temperatures \cite{seo08}. There have been many studies aiming to construct analytical models that contain impurities. Most of these models assume that the impurities are present in different concentrations on the catalytic surface. In this work we analyze the problem in a more realistic way by assuming that the impurities are part of the gas mixture \cite{bust00, schm01, hua03}. The impurities can be adsorbed or desorbed, but otherwise they do not take part in the reaction. Desorption of CO is included as a way to mimic temperature effects. In order to be able to study the effects of a nonzero CO desorption rate in isolation, we 
ignore other temperature dependent effects, such as lateral diffusion of CO molecules, or competition between CO 
desorption and oxidation. 

In a previous work \cite{buen12} that did not include CO desorption, we found that, if the impurities do not desorb, 
the reactive window that characterizes the ZGB model disappears. If they can desorb, a reactive window reappears if the proportion of contaminants does not surpass a critical value. This critical value depends on the desorption rate of the impurities. Continuing this line of work, 
in the present article, besides impurities, we include the well documented process of CO desorption, discussed above. 

The rest of this paper is organized as follows. In Sec.\ II we describe how we modify the ZGB model to eliminate the unphysical continuous phase transition at $y_1$, to include the presence of impurities in the gas phase, and to include a CO desorption rate. In Sec.\ III we present our numerical results for the modified model. We present the results in two 
separate stages. First we study a system in which the impurities remain on the surface after being adsorbed 
(Sec.\ III~A), and next 
we analyze the case that they can also desorb (Sec.\ III~B). In Sec.\ IV we present our conclusions and topics 
for future research.

\section{Model and Simulations}

We study the catalytic oxidation of CO on a surface in contact with a gas phase that consists of a mixture of CO, O$_2$, and impurities X, in different proportions. The impurities can be adsorbed on the surface, where they do not react with the other adsorbates. Once on the surface, the X and the CO can be desorbed at different rates. We further modify the ZGB model by changing the adsorption mechanism of the O$_2$ molecule, so that it is adsorbed on two next-nearest-neighbor (nnn) vacant sites (separated by $\sqrt 2$ lattice units) instead of two nearest-neighbor (nn) sites. It has 
previously been shown that this minor change eliminates the unphysical O-poisoned phase \cite{ojed12,buen12}. 

The model  is simulated on a square lattice of
linear size $L$ that represents the catalytic surface. A Monte Carlo simulation generates a sequence of trials: CO, O$_2$, or X adsorption, or X or CO desorption. A site $i$ is chosen at random.  If it is occupied by a CO or a X, we attempt desorption with probabilities $k_c$ and $k_x$, respectively. If the site is empty, we
attempt adsorption: CO with probability $y$, X with probability $y_x$, or  O$_2$ with probability $1-y-y_x$. These probabilities are proportional to the amounts of the different species in the gas phase. The O$_2$ molecule can only be adsorbed on a pair of vacant nnn sites. In a O$_2$ adsorption attempt a nnn of site $i$ is selected at random; if it is occupied the trial ends, if not the adsorption proceeds and the O$_2$ molecule is adsorbed and dissociates into two O atoms. After a CO or O$_2$ adsorption event is realized, all nn pairs are checked in random order. Pairs consisting of 
nn CO and O atoms react: a CO$_2$ molecule is released, and two nn sites are vacated. 
A schematic representation of this
algorithm is given by the reaction equations,
\begin{eqnarray}
\text{CO(g)} + \text{S} & \rightarrow & \text{CO(a)}
\nonumber \\
\text{O}_2 + 2\text{S} & \rightarrow & 2\text{O(a)}
\nonumber \\
\text{CO(a)} + \text{O(a)} & \rightarrow & \text{CO}_2\text{(g)} + 2\text{S}
\\
\text{X(g)} + \text{S} & \rightarrow & \text{X(a)}
 \nonumber \\
\text{X(a)} &\rightarrow& \text{X(g)}+ \text{S}
\nonumber\\ 
\text{CO(a)} &\rightarrow& \text{CO(g)}+ \text{S}
\nonumber
\;
\nonumber
\end{eqnarray}
Here S represents an empty site on the surface, $\rm g$ means gas phase, and $\rm a$ means adsorbed.
The first three steps correspond to a Langmuir-Hinshelwood mechanism. This study differs from our previous one~\cite{buen12} in the presence of the last reaction, which simulates the CO desorption.

The simulations are performed on a square lattice of $L\times L$ sites, with $L=120$ and periodic boundary conditions. 
The time unit is one Monte Carlo Step per Site, MCSS, in which each
lattice site is visited once, on average. Averages are subsequently taken over $10^5$
MCSS after $10^4$ MCSS are used to achieve a stationary state. 

Coverages of the different species are defined as the fractions of sites on the surface occupied by the species. 
We calculate the CO, O, and X coverages, and the rate of production of CO$_2$. Always starting from an empty lattice, 
the system was allowed to reach a stationary state before data were recorded for analysis.  
(Any initial coverages of irreversibly adsorbed species would obviously remain on the surface indefinitely, 
thus changing the resulting stationary state.)

\section{Results} 

We start by analyzing the case in which the impurities do not desorb in Sec. III~A. Then we study the effects of 
impurity desorption in Sec. III~B.

\subsection{ No desorption of impurities, $k_x=0$}
In the case that the impurities do not desorb, the fifth step in the reaction described by Eq.~(1) does not occur, and the impurities, once adsorbed, remain on the surface. This case can be useful to understand the behavior of some catalytic systems that become irreversibly poisoned by impurities. 
The behavior of a system with this type of impurities is shown in Fig.~1, where we plot the coverages vs $y$ for a system with a fixed proportion of impurities in the gas phase,  for different values of  $k_{c}$. The first difference from the standard ZGB-model to notice, is the absence of the O poisoned phase. There is no continuous transition at $y_1$: the CO coverage departs from zero as soon as $y > 0$. As discussed above, this is fully explained by the change in the adsorption mechanism of O and does not depend on CO desorption \cite{buen12}. 

Not surprisingly, when impurities that do not 
desorb are added to the gas phase, the reaction rate vanishes in the steady state, independently of the value of $k_{c}$.
 Consequently, the discontinuous transition to a separate, poisoned state disappears. The CO and O coverages change smoothly between their extreme values, as seen in Fig.~1(a) and Fig.~1(b), respectively. Previous results for the case $k_{c}=0$ indicate that this is also true for any value of $y_x \ne 0$ \cite{buen12}. As the CO desorption rate 
increases, the CO coverage decreases gradually until, for sufficiently high values of $k_{c}$, the surface does not contain any CO, Fig.~1(a). As expected, the O coverage decreases as $y$ increases, and it has a relatively weak
 dependence on $k_{c}$, increasing slowly as $k_{c}$ increases, Fig.~1(b). The behavior of the X coverage strongly depends on $k_c$. For a
relatively low $k_{c}$, it reaches a maximum at a value of $y$ that depends on $k_{c}$ (and $y_x$), and then decreases. But for larger values of $k_{c}$, it increases monotonically with $y$ until the surface is almost completely filled with 
 impurities, Fig.~1(c). 

In a previous work \cite{buen12} we analyzed the case in which there was no CO desorption 
(curves corresponding to  $k_{c}=0$ in Fig.~1). In the present work we find that, when a nonzero 
CO desorption rate is included, the impurities take the empty spaces left by the CO. For sufficiently large values of $k_{c}$, the surface becomes almost totally covered by X (and a few O), while in the former case ($k_c=0$) the surface always ends up filled with CO. 
Since O adsorption requires two empty sites, the oxygen coverage is only weakly enhanced by the CO desorption. 
For intermediate values of $k_c$ and values of $y$ beyond approximately 0.5, the steady state is a mixture of CO and X, 
almost entirely devoid of O.

\subsection{Non-zero impurity desorption, $k_{\rm x} > 0$}
Next we study the changes that occur when there is a nonzero probability that an adsorbed impurity can leave the surface, i.e., when the fifth step of Eq.~(1) is included. Now both CO and X can be desorbed from the surface with probabilities 
$k_c$ and  $k_{x}$, respectively. We fix the proportion of X in the mixture, $y_x$, and the CO desorption rate, $k_c$, 
then vary $k_x$. We present the coverages vs $y$ for different values of $k_{x}$ for a small value of $k_{c}$ ($k_{c}=0.005$) in Fig.~2, and for a larger value ($k_{c}=0.03$, comparable to the critical value of $k_c \approx 0.04$ in the impurity-free ZGB model) in Fig.~3. For comparison, the figures also include the case in which there are no impurities and the CO can desorb, $y_x=0$ and $k_c > 0$, and the case, already discussed, in which $k_{x}=0$. 
In the previous section we showed that, if the impurities cannot desorb, the productive phase, and consequently also the discontinuous transition at $y_2$ to a poisoned phase disappear, independently of the value of $k_{c}$. 
In this section we show that a nonzero rate of CO$_2$ production reappears for $k_x > 0$.
As can be seen in Figs.~2 and 3 for small values of $k_{x}$, the coverages change continuously between their extreme values. When $k_{x}$ reaches a sufficiently large value, which depends on $k_{c}$, a discontinuous transition to a 
fully poisoned phase also reappears. 
Not surprisingly, for large values of $k_x$ the model becomes almost indistinguishable from the model without impurities. 
We find that the CO, O, and X coverages depend strongly on $k_x$, as seen in Figs.~2 and 3. The X coverage, Fig.~2(c) and Fig.~3(c), reaches a maximum value that decreases as $k_x$ increases. The location of the maximum moves toward higher values of $y$. For sufficiently high values of $k_x$, there are no impurities on the surface in the steady state for any value of $y$.  In Fig.~4 we present the reaction rate vs $y$ for the values of $y_x$, $k_c$, and $k_x$ analyzed in 
Figs.~2 and 3. For comparison, the figures also include the case in which there are no impurities (curves labeled $y_x=0$). As in the case without CO 
desorption \cite{buen12}, the presence of impurities reduces the maximum reaction rate and shrinks the reactive window. For a fixed value of $y_x$, the rate of production of CO$_2$ increases with increasing $k_x$ until, for a sufficiently large value of $k_x$, it reaches approximately the same values as in the system without impurities. At high values of $k_x$ the impurities do not play any role in the steady state, and the system is almost indistinguishable from the one without impurities.

To further understand the effect of including a CO desorption rate in the system with impurities, in Fig.~5 we compare the CO$_2$ production rate vs $y$ for systems with  different values of $k_{ c}$ at the same partial pressure of X ($y_x=0.005$) and same $k_{x}$ ($k_{ x}=0.001$). As seen from Fig.~4, at these values of $y_x$ and $k_x$, 
the CO$_2$ production is generally quite depressed. For the values analyzed, when $k_{c}=0$ the CO$_2$ production increases until it decreases sharply to zero at $y_2\approx 0.25$. 
(This value of $y_2$ depends strongly on $y_x$, as shown in Fig.~5 of Ref.~\onlinecite{buen12}.)
When $k_{c}$ is included, the discontinuity disappears, the production rate reaches a maximum and then decreases smoothly toward zero. For values of $y$ below the point where the maximum CO$_2$ production is reached, the production depends very little on $k_{c}$. As $k_{c}$ increases, the maximum value of the production rate increases slowly until it reaches a limiting value that seems to be independent of $k_{c}$. It is important to notice that, even if the maximum value of the CO$_2$ production reaches a limiting value, the size of the reactive window increases with further
increasing values of $k_{c}$. At large values of $k_{c}$ the system produces CO$_2$ almost in the entire range of values of $y$.

Taken together, the coverage and reaction-rate data shown in Figs.~2-5 suggest the possibility that there is a range 
of intermediate values of $k_x$ and $k_c$, for which a {\it continuous\/} transition to a poisoned phase 
might occur near $y = 0.5$. 
However, confirmation or negation of this possibility will require much more detailed simulations, including the use of 
finite-size scaling techniques to analyze data obtained from different system sizes. We reserve this for a future study.

\section{Conclusions}
In this work we study a model of the reaction CO+O$_2$ on a catalytic surface in contact with a gas phase that 
contains CO, O$_2$, and impurities, X. The impurities can be adsorbed on the surface, where they do not react with 
the other adsorbates.  Once adsorbed, X and CO particles can desorb with probabilities $k_{ x}$ and $k_{c}$, respectively. Our model is based on a ZGB model that has been modified such that the two oxygen atoms adsorb at two nnn sites 
instead of two nn sites. This modification removes the unphysical oxygen-poisoned phase predicted by the original ZGB 
model. By comparing this study with a recent work that included reversibly adsorbed impurities,
 but excluded CO desorption ($k_{c}=0$) \cite{buen12}, we provide a possible mechanism for the well documented 
 temperature effects (commonly modeled by the introduction of desorption rates) in catalytic systems. 

In order to study the effects of CO desorption in isolation, we deliberately exclude other temperature dependent effects, such as CO diffusion or a competition between CO desorption and oxidation. Since the concentration of vacancies in the adsorbate layer is quite low under all the conditions we have investigated, we do not believe the former should have much effect. As for the latter, we believe it would probably simply have an effect similar to a reduction of the CO pressure $y$. 

The first effect of including CO desorption ($k_{c} > 0$) that we observe, is that, for sufficiently high values of $k_c$ and $y$, once the steady state has been reached the surface is almost completely filled with X. In contrast, when $k_{c}=0$, at large values of $y$, for the same value of $y_{x}$, the surface ends up almost completely covered by CO \cite{buen12}. This effect is observed independently of whether there is impurity desorption or not. 

If the impurities do not desorb ($k_{x}=0$),  the system does not produce CO$_2$ in the steady state, 
and the characteristic discontinuous transition to a poisoned phase disappears. This effect is not affected by the 
presence of CO desorption $k_{c} > 0$. However when the impurities are allowed to desorb ($k_{x} > 0$), the 
CO$_2$ production rate reappears. The discontinuous transition at $y_2$ to a poisoned phase also reappears at 
sufficiently high values of $k_{x}$. When CO desorption is included, the coverage changes become smoother, 
and the discontinuity only reappears at much higher values of $k_x$.  For a fixed value of $k_{c}$ and $y_{x}$, $y_2$ increases with $k_x$ until it reaches the same value of $y_2$ as in the system without impurities.  
Our data also suggest the possibility that there 
might exist a region in the $k_x, k_c$ space, in which the discontinuous transition to a poisoned phase becomes continuous. 
Exploration of this possibility is reserved for a future study. 

The most dramatic effect of the existence of CO desorption can be observed in the CO$_2$ 
production rate of the system. Its maximum value increases slowly with increasing values of $k_c$, but most importantly, the reactive window widens in such a way that now the system remains catalytically active in the whole range of values of $y$. This is a very interesting result that can explain why systems with impurities recover their catalytic activity when the temperature increases. The observation that a high enough CO desorption rate can somewhat reduce the negative effects of the impurities is worth further study.

\section*{Acknowledgments}
G.M.B is grateful for the hospitality of the Physics Department at Florida State University. P.A.R acknowledges support by U.S. National Science Foundation Grant No. DMR-1104829.

\begin{figure}
\includegraphics[scale=.34]{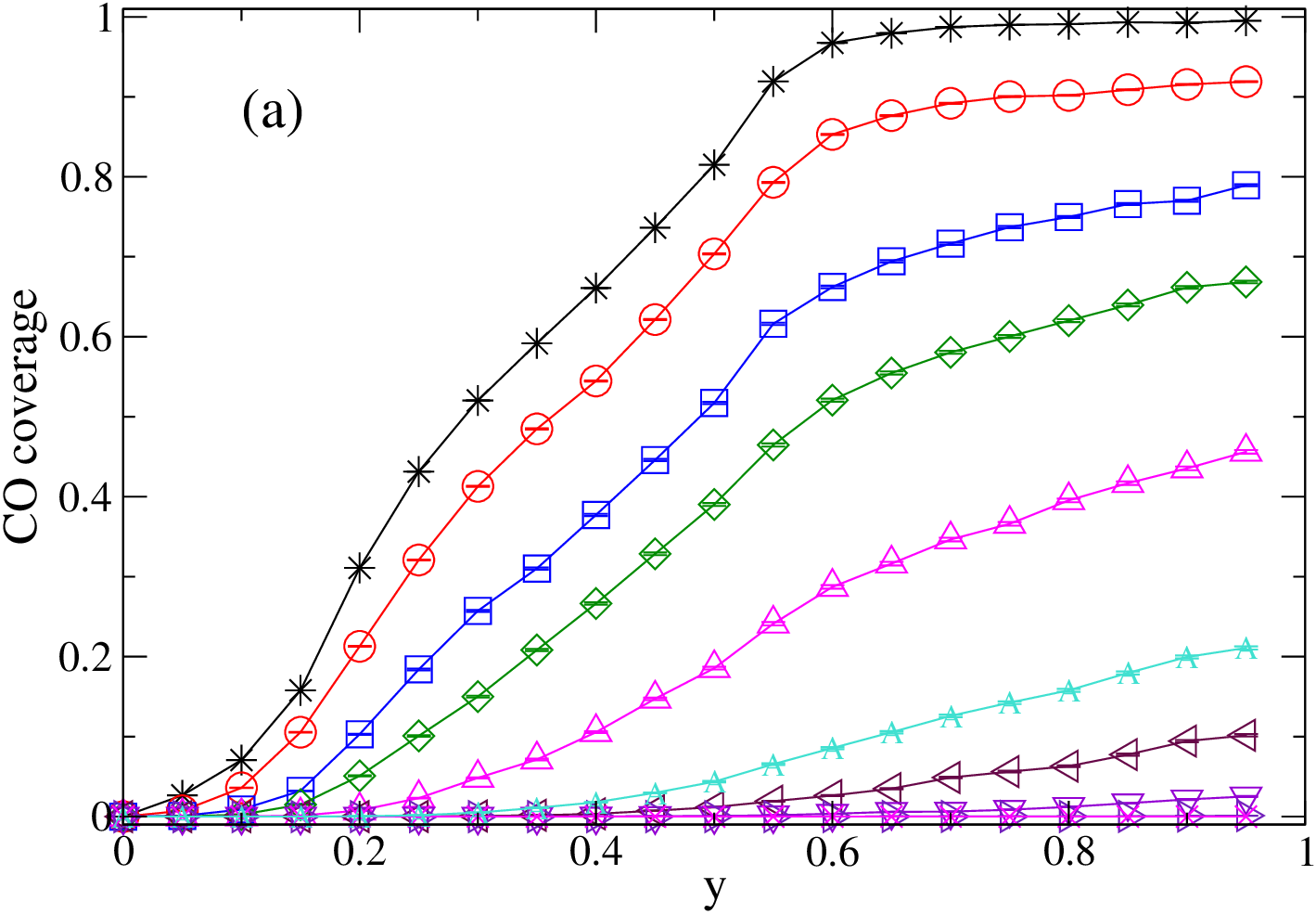}\\
\vspace{1.0truecm}
\includegraphics[scale=.34]{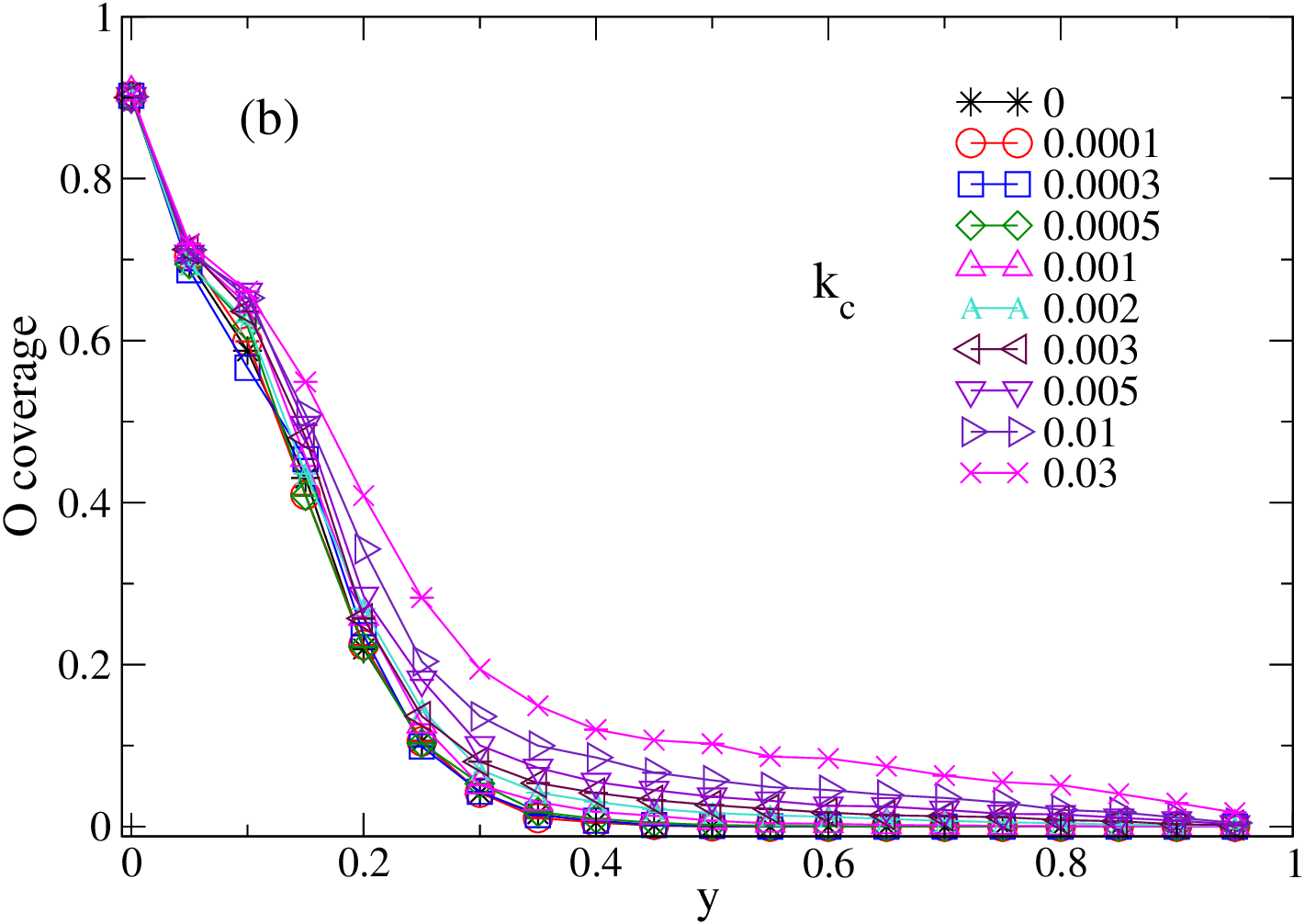}\\
\vspace{1.0truecm}
\includegraphics[scale=.34]{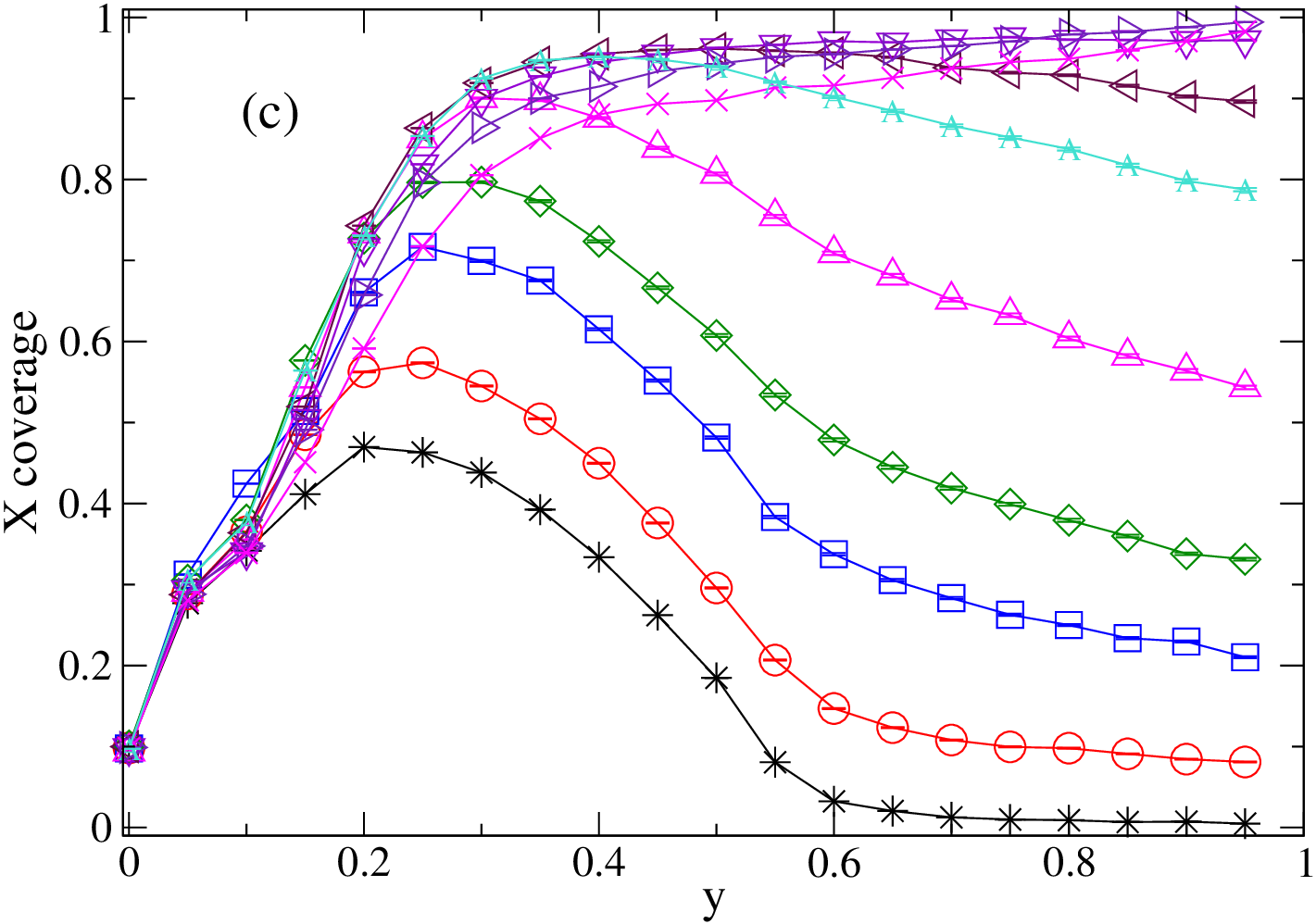}\\
\caption[]{(Color online) Coverages vs $y$ for the case that there is no impurity desorption, $k_x=0$, for fixed 
$y_x=0.005$, and the values of $k_{c}$ indicated in part (b). (a) CO coverage, (b) O coverage, (c) X coverage.  
The results for $k_c=0$ ($*$) correspond to those represented by squares in 
Fig.~1 of Ref.~\protect\onlinecite{buen12}.
We did not plot the reaction rates because they are practically zero for all $k_c$.} 
\label{f1}
\end{figure}

\begin{figure}
\centering\includegraphics[scale=.36]{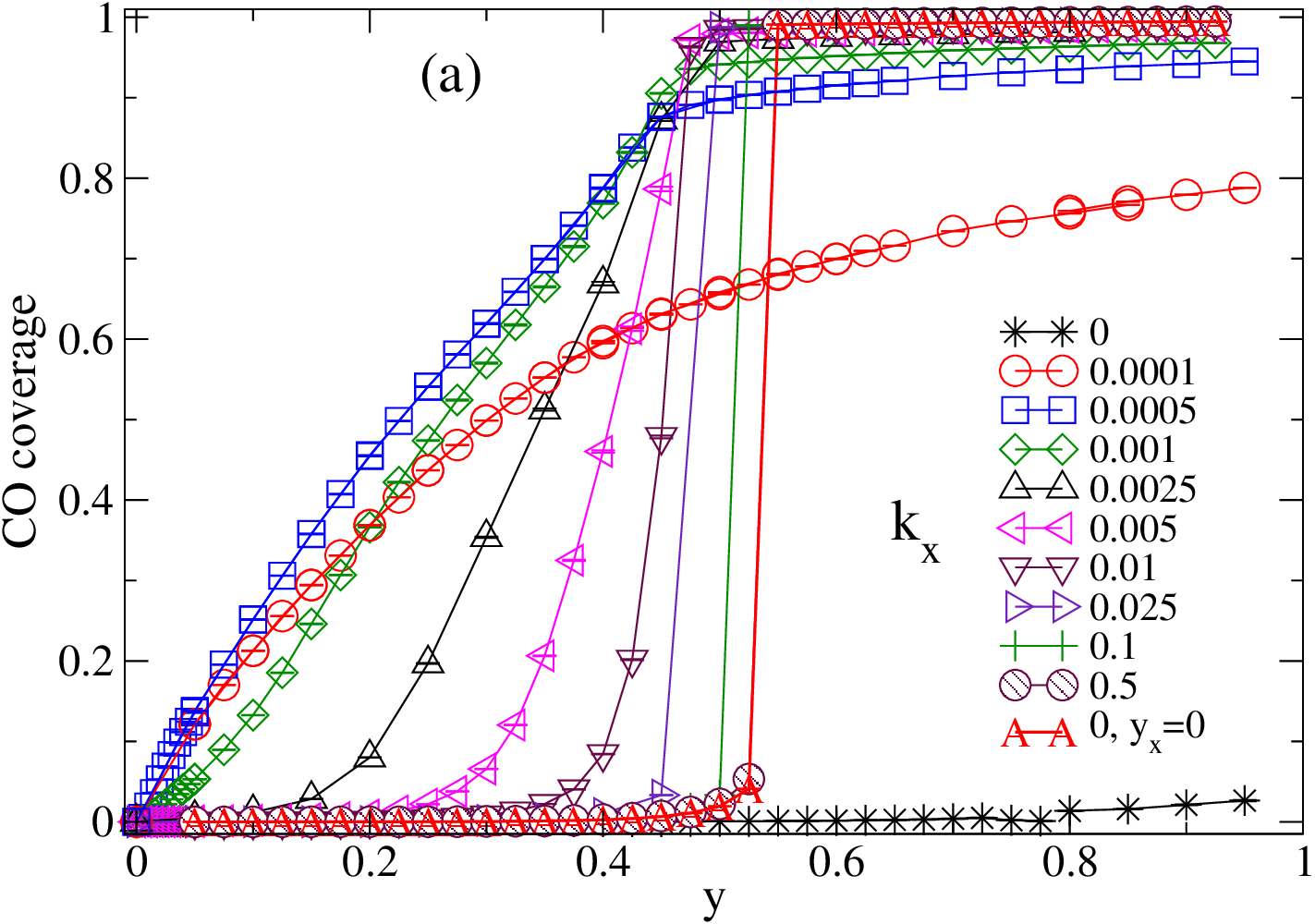}\\
\vspace{1.2truecm}
\centering\includegraphics[scale=.36]{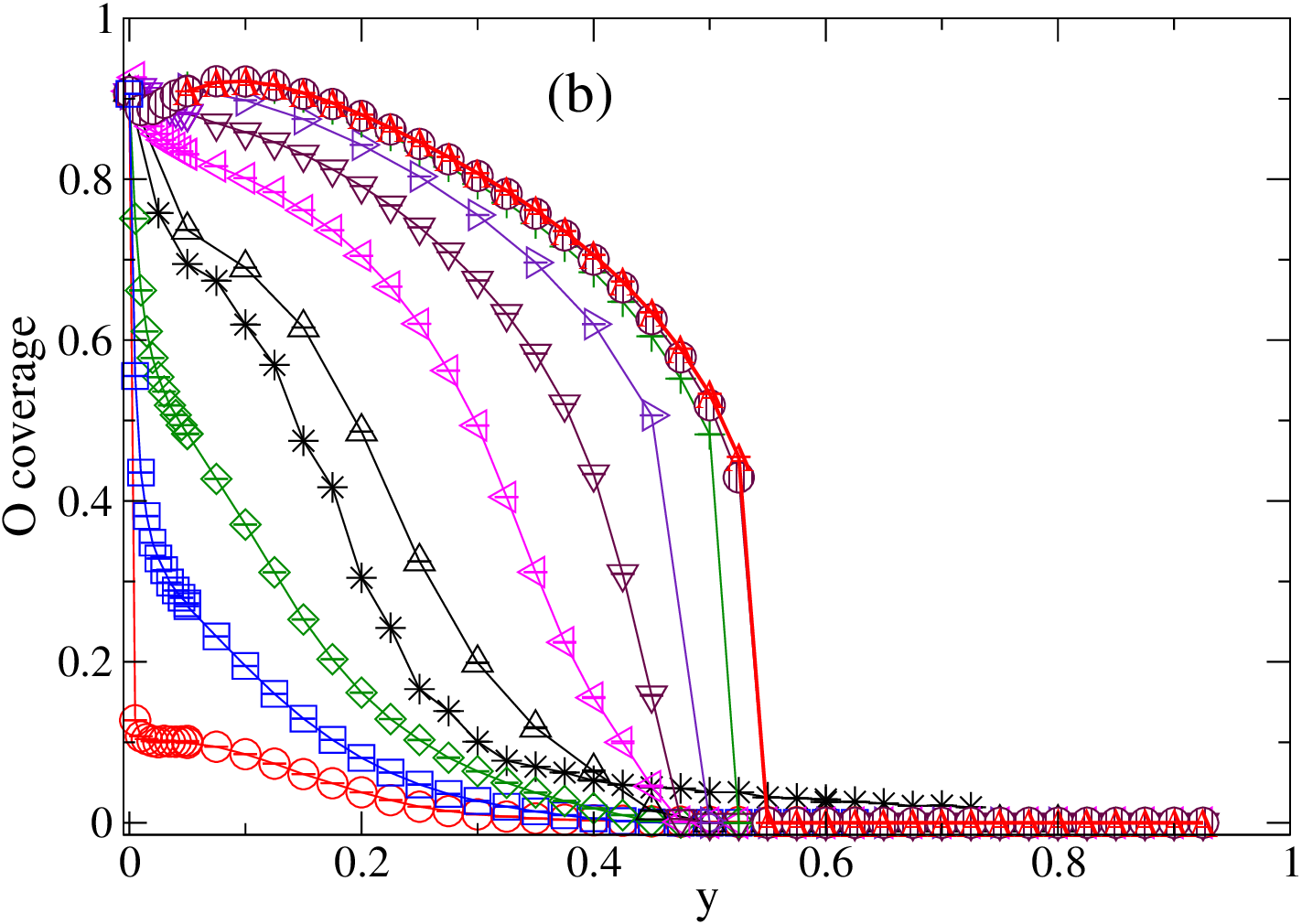}\\
\vspace{1.2truecm}
\centering\includegraphics[scale=.36]{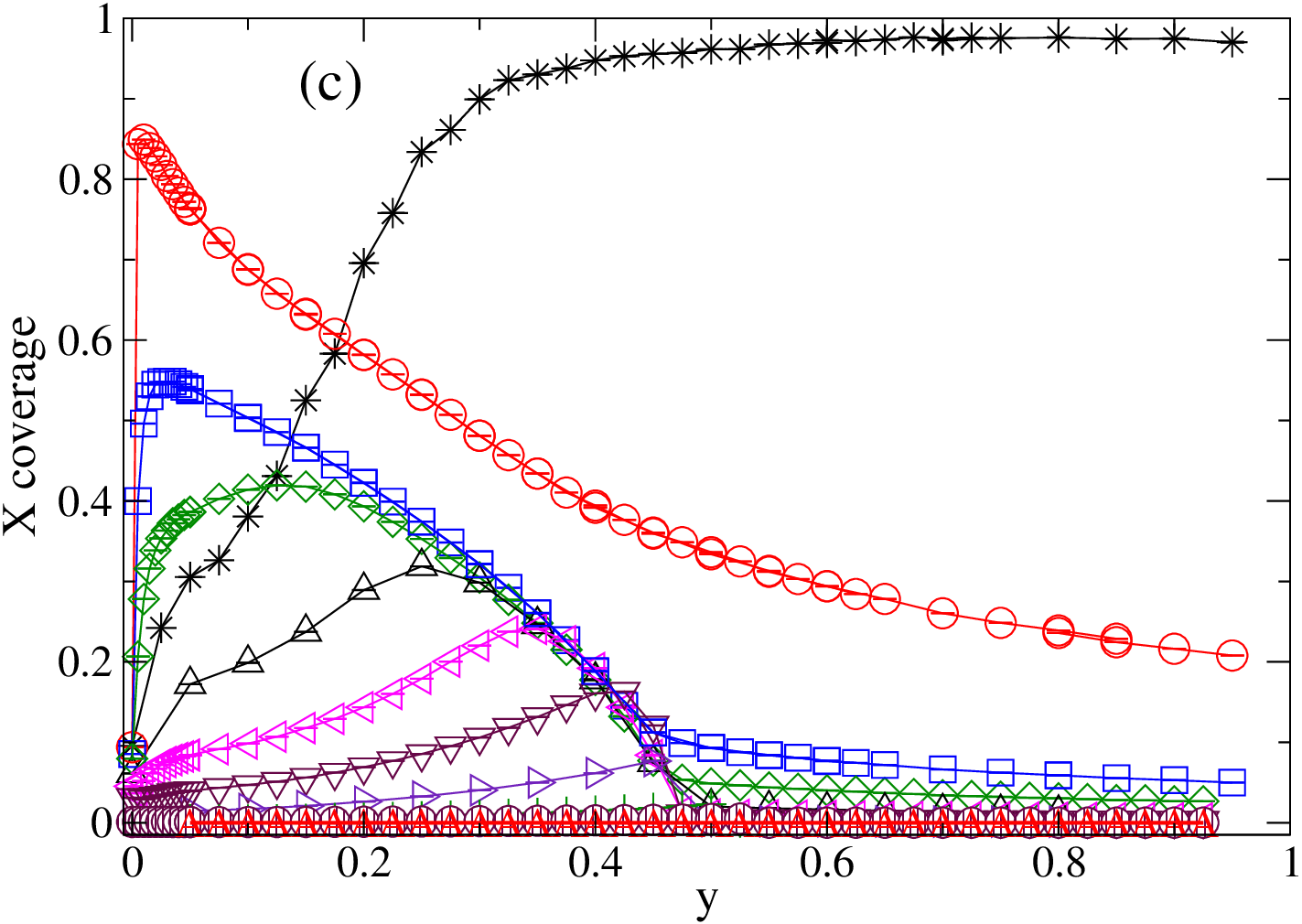}\\
\vspace{0.4truecm}
\caption[]{(Color online) Coverages vs $y$ when $k_{c}=0.005$ for fixed $y_x=0.005$ and different values of $k_x$. (a) CO coverage (b) O coverage (c) X coverage. The values of $k_x$ are indicated in part (a).}
\label{f2}
\end{figure}

\vspace{2.truecm}

\begin{figure}
\centering\includegraphics[scale=.36]{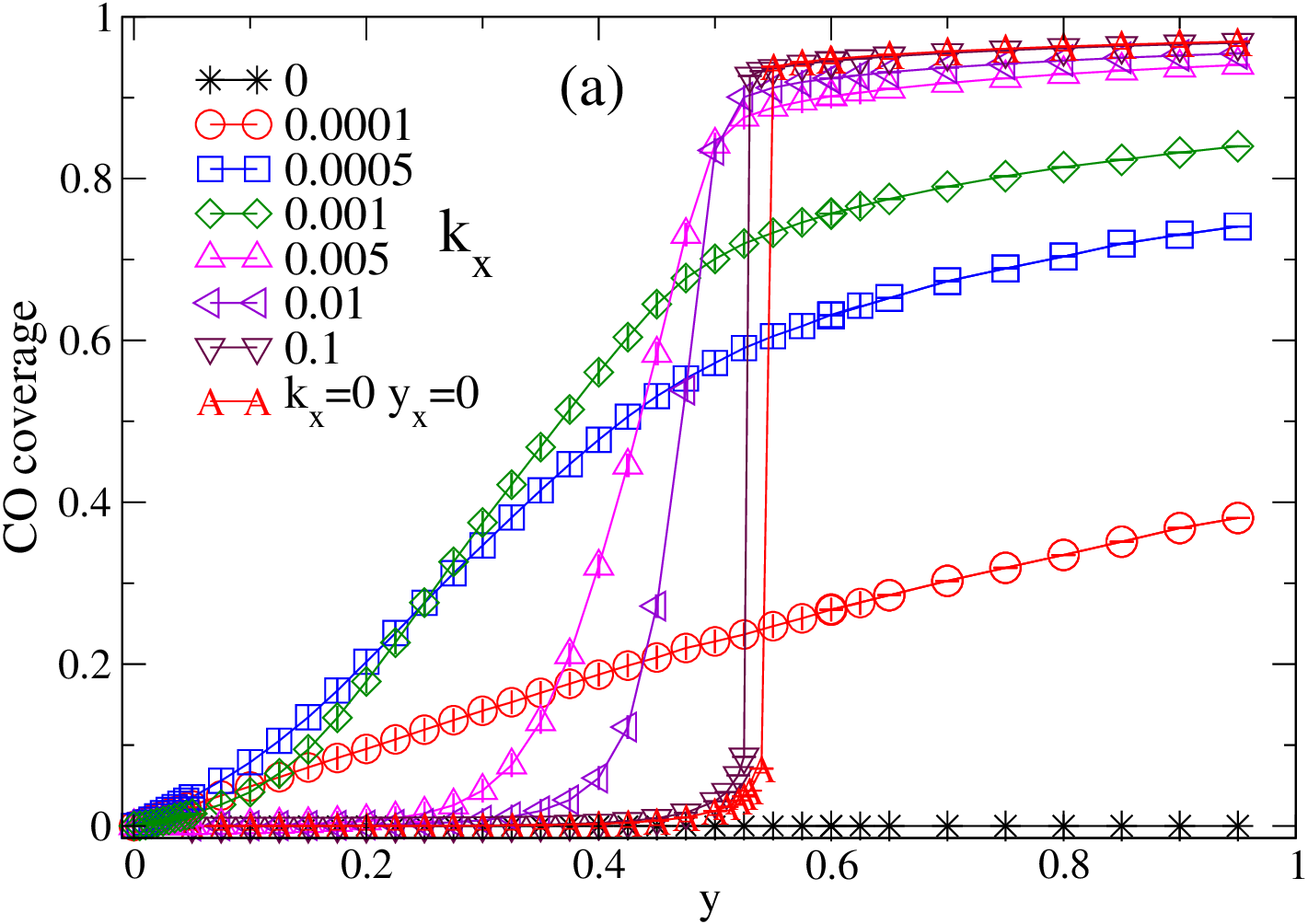}\\
\vspace{1.2truecm}
\centering\includegraphics[scale=.36]{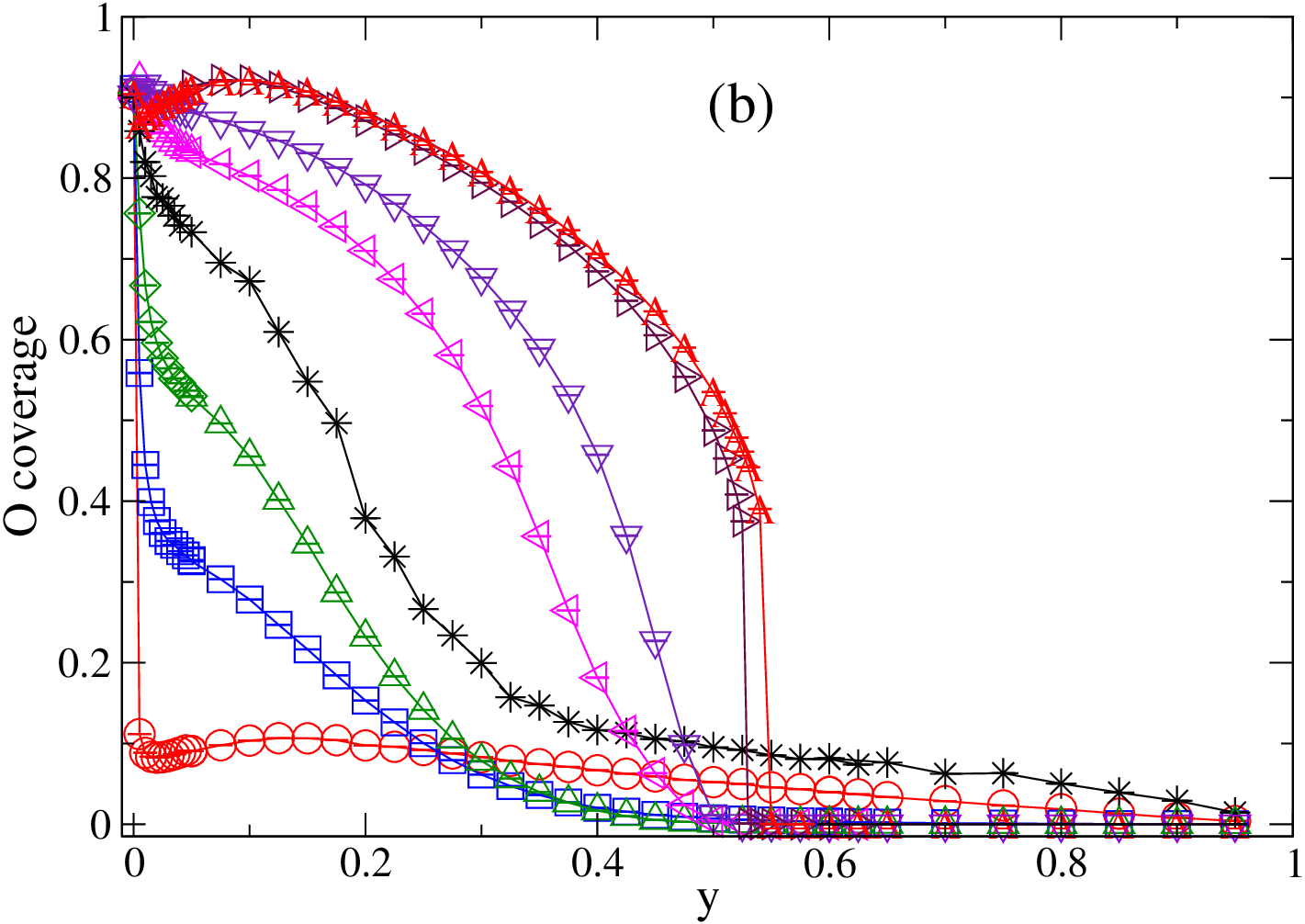}\\
\vspace{1.2truecm}
\centering\includegraphics[scale=.36]{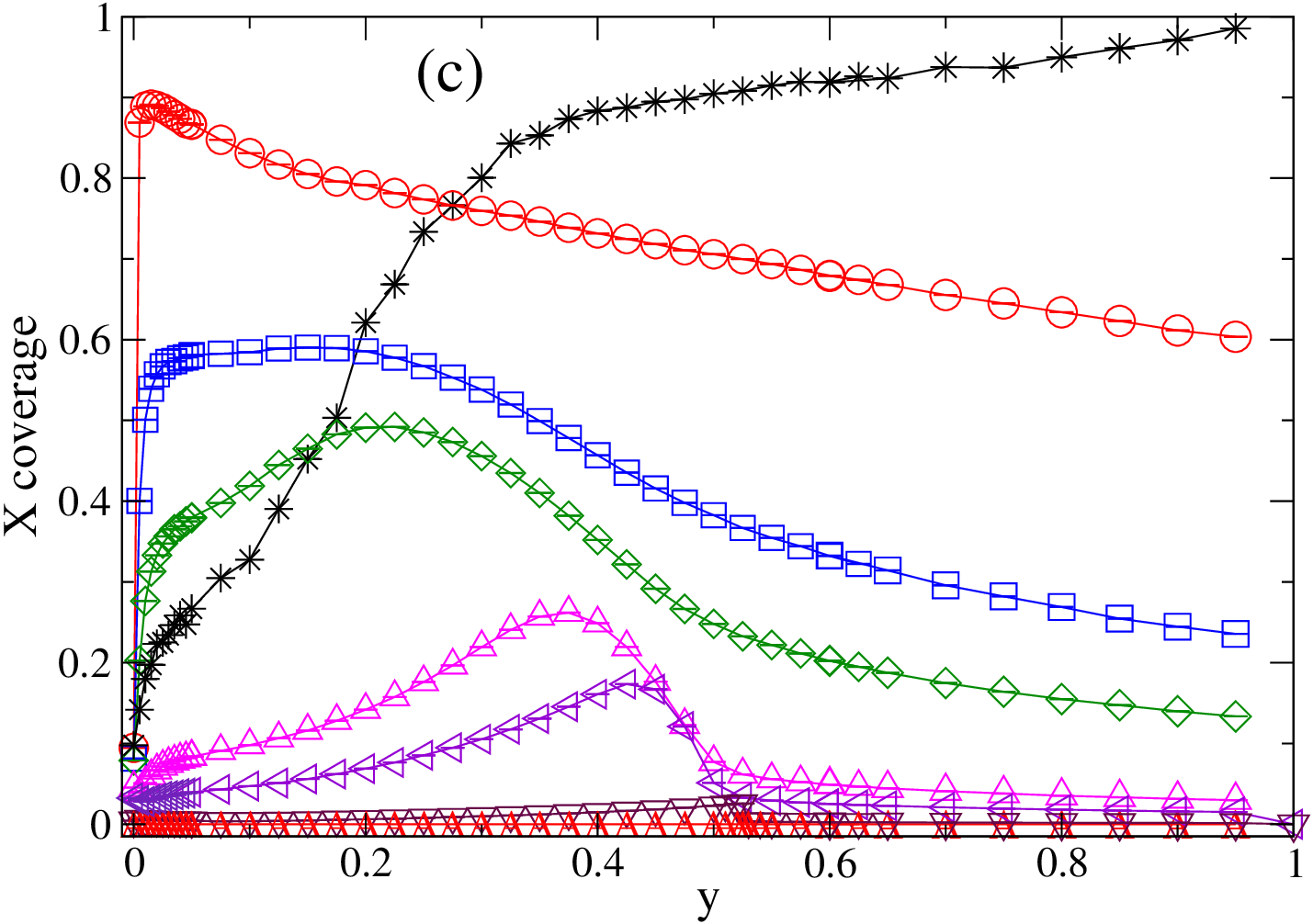}\\
\vspace{0.4truecm}
\caption[]{(Color online) Coverages vs $y$ when $k_{c}=0.03$ for fixed $y_x=0.005$ and different values of $k_x$ indicated in (a). (a) CO coverage (b) O coverage (c) X coverage. }
\label{f3}
\end{figure}

\begin{figure}
\centering\includegraphics[scale=.38]{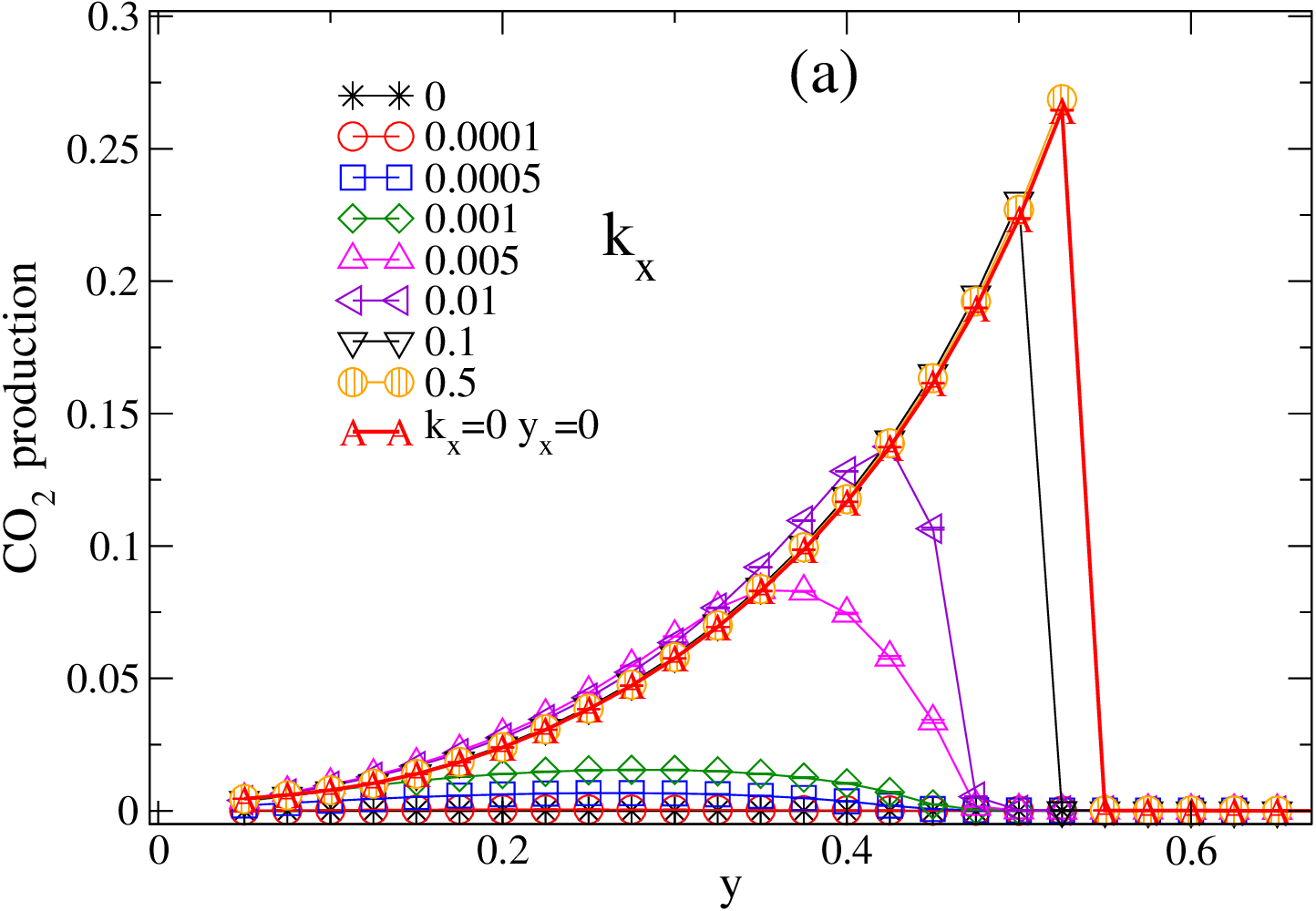}\\
\vspace{1.2truecm}
\centering\includegraphics[scale=.38]{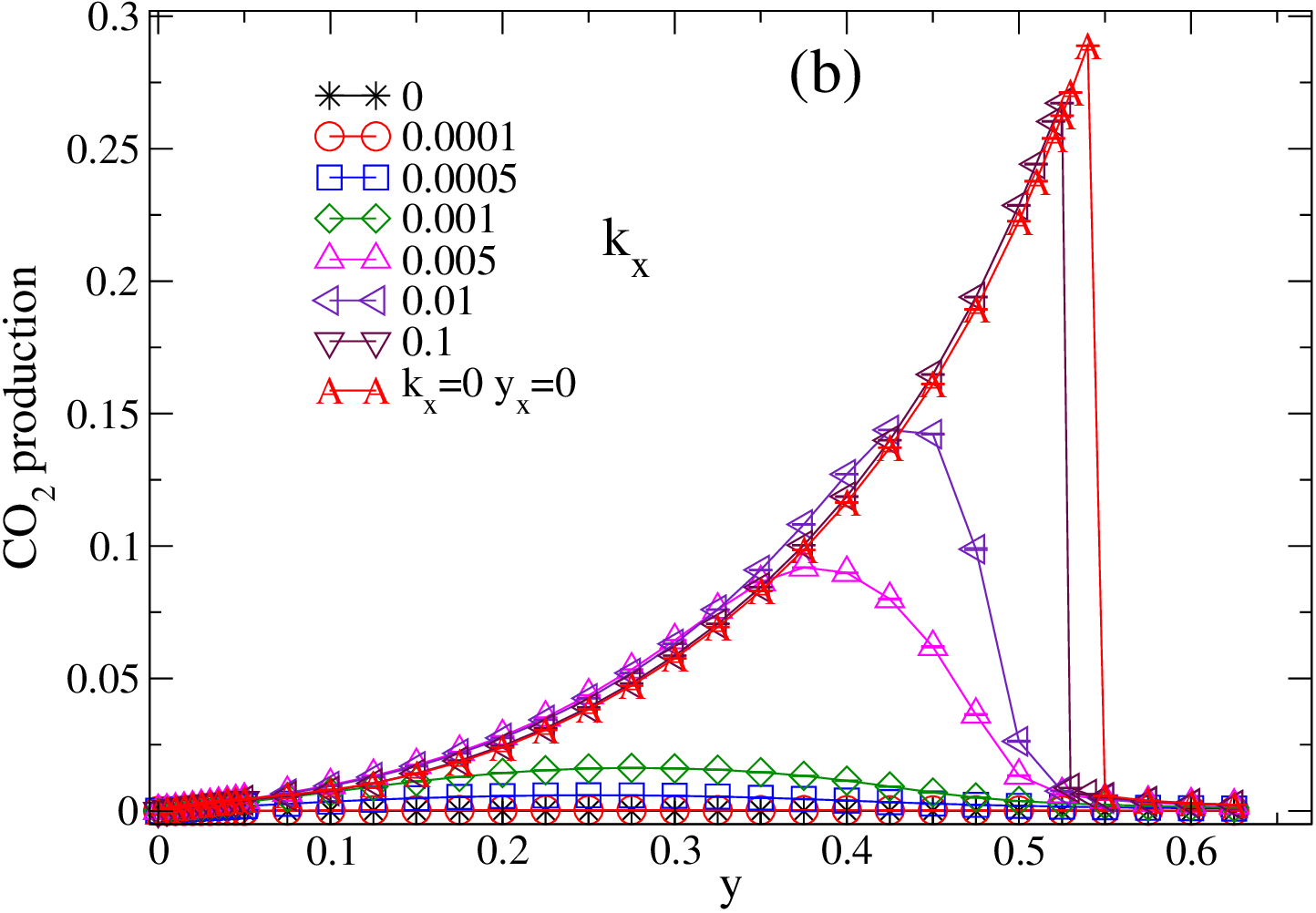}\\
\vspace{1.2truecm}
\vspace{1.2truecm}
\caption[]{(Color online) Production rate of CO$_2$  vs $y$ for fixed $y_x=0.005$ and different values of $k_x$. (a) $k_{\rm c}=0.005$ (b) $k_{\rm c}=0.03$.}
\label{f4}
\end{figure}

\begin{figure}
\centering\includegraphics[scale=.46]{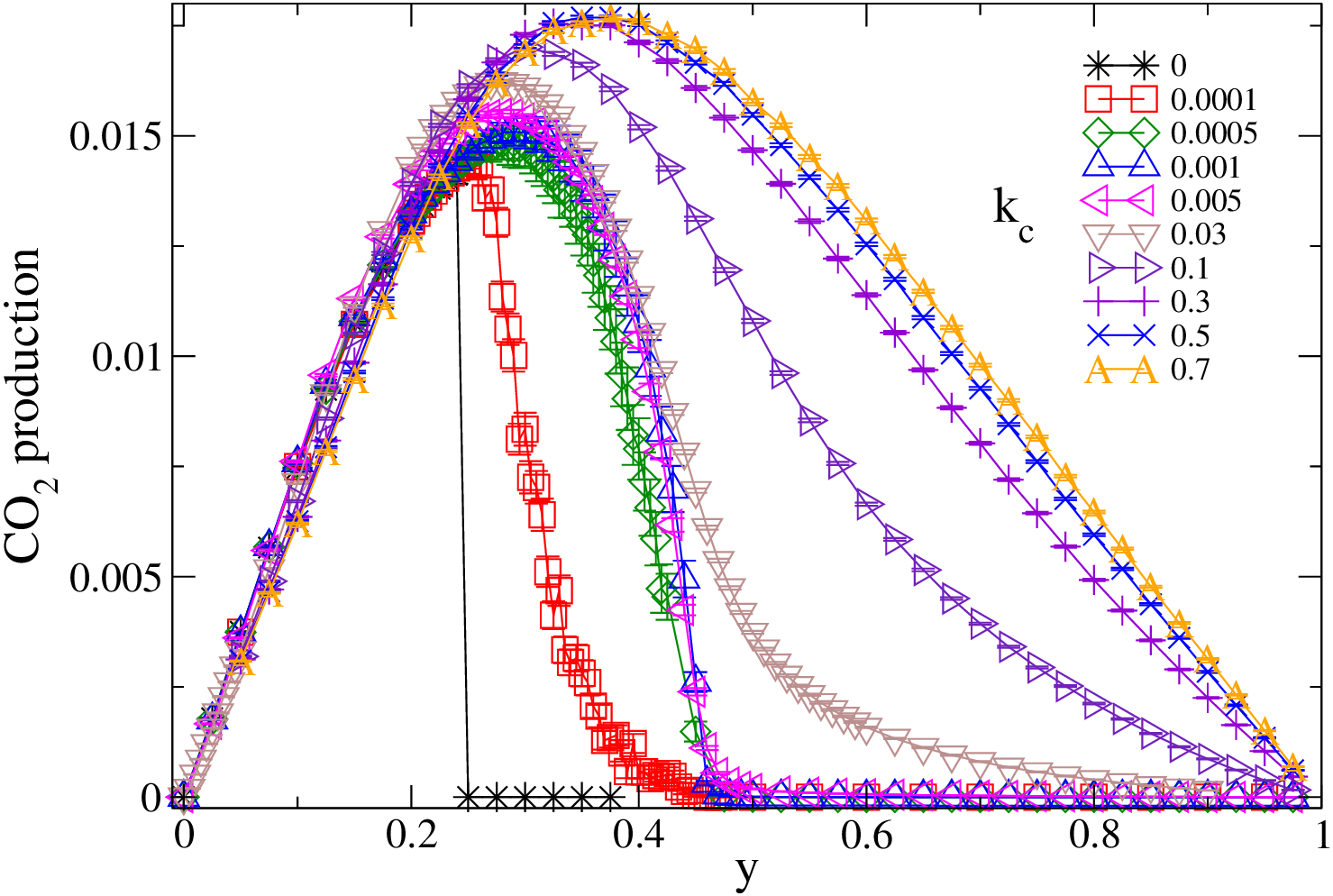}\\
\caption[] {(Color online) Production rate of CO$_2$ vs $y$ for different values of $k_c$ at fixed $y_x=0.005$  and $k_x=0.001$. }
\label{f5}
\end{figure}


\begin{thebibliography}{99}

\bibitem{baxt02} 
R.~J.~Baxter, P.~Hu, {J.~Chem.~Phys.} \textbf {116}, 4379 (2002).


\bibitem{chri94}K.~Christmann,
\textit{Introduction to Surface Physical Chemistry}, Steinkopff
Verlag, Darmstadt, 1991; V.~P.~Z.\ Zhdanov and B.~Kazemo, Surf. Sci.
Rep. \textbf {20}, 111 (1994).

\bibitem{yama95} S.~Y.~Yamamoto, C.~M.~Surko, M.~B.~Maple, and R.~K.~Pina, {J.~Chem.~Phys.}
\textbf{102}, 8614 (1995).

\bibitem{imbi95} R.~Imbihl and G.~Ertl, {Chem.~Rev.} \textbf{95}, 697 (1995).

\bibitem{marro99} J.~Marro and R.~Dickman,
\textit{Nonequilibrium Phase Transitions in Lattice Models},
Cambridge University Press, 1999, ch. 5, pp. 141-160.

\bibitem{herz81} R.~K.~Herz, {Ind.\ Eng.\ Chem.\ Prod.\ Res.\ Dev.} \textbf{20}, 451 (1981).

\bibitem{liu93} D.~R.~Liu and J.~S.~Park, {App.\ Catal.\ B} \textbf{2}, 49 (1993).

\bibitem{asak97} K.~Asakura, J.~Lauterbach, H.~H.~Rotermund, and G.~Ertl, {Surf.\ Sci.}  \textbf{374}, 125 (1997).

\bibitem{forz99} P.~Forzatti and L.~Lietti, {Catal.\ Today} \textbf{52}, 165 (1999).

\bibitem{yu98} T.-C.~Yu and H.~Shaw, {App.\ Catal.\ B} \textbf{18}, 104 (1998).

\bibitem{beck95} D.~D.~Beck and J.~W.~Sommers, {App.\ Catal.\ B} \textbf{6} , 185 (1995).

\bibitem{buen12}  G.~M.~Buend\y a and P.~A.~Rikvold,  {Phys.\ Rev.\ E}
\textbf{85}, 031143 (2012).

\bibitem{EIGE78}
G.~Eigenberger, Chem.\ Eng.\ Sci.\ {\bf 33}, 1263 (1978).

\bibitem{VIGI96}
R.~D.\ Vigil and F.~T.\ Willmore, Phys.\ Rev.\ E {\bf 54}, 1225 (1996). 

\bibitem{engl78} T.~Engl and G.~Ertl, {J.\ Chem.\ Phys.}
\textbf{69}, 1267 (1978).

\bibitem{ziff86} R.~M.\ Ziff, E.~Gulari, and Y.~Barshad, {Phys.\ Rev.\ Lett.}
\textbf{56}, 2553 (1986).

\bibitem{VOLK01}
S.~V{\"o}lkening and J.~Wintterlin, J.\ Chem.\ Phys.\ {\bf 114}, 6382 (2001).

\bibitem{PETR05}
N.~V.\ Petrova and I.~N.\ Yakovkin, Surf.\ Sci.\ {\bf 578}, 162 (2005). 

\bibitem{LIU06}
D.-J.\ Liu and J.~W.\ Evans, J.\ Chem.\ Phys.\ {\bf 124}, 154705 (2006).

\bibitem{NAGA07}
M.~Nagasaka, H.~Kondoh, I.~Nakai, and T.~Ohta, J.\ Chem.\ Phys.\ {\bf 126}, 044704 (2007).

\bibitem{ROGA08}
J.~Rogal, K.~Reuter, and M.~Scheffler, Phys.\ Rev.\ B {\bf 77}, 155410 (2008).

\bibitem{LIU09}
D.-J.\ Liu and J.~W.\ Evans, Surf.\ Sci.\ {\bf 603}, 1706 (2009).

\bibitem{HESS12}
F.~Hess, A.~Farkas, A.~P.\ Seitsonen, and H.~Over, J.\ Comput.\ Chem.\ {\bf 33}, 757 (2012).

\bibitem{kauk89} H.~P.\ Kaukonen and R.~M.\ Nieminen, {J.\ Chem.\ Phys.}
\textbf{91}, 4380 (1989).

\bibitem{bros92} B.~J.~Brosilow and R.~M.\ Ziff, {Phys.\ Rev.\ A}
\textbf{46}, 4534 (1992).

\bibitem{alba92} E.~V.~Albano, {Appl.\ Phys.\ A: Solids Surf.} \textbf{54}, 2149 (1992).

\bibitem{meak90} P.~Meakin, {J.\ Chem.\ Phys.} \textbf{93}, 2903 (1990).

\bibitem{buen09}
G.~M.~Buend\y a, E.~Machado. and P.~A.~Rikvold, {J.\ Chem.\ Phys.}, \textbf{131}, 184704 (2009).

\bibitem{ehsa89} M.~Ehsasi, M.~Matloch, O.~Frank, J.~H.\ Block, K.~Christmann,
F.~S.\ Drys, and W.~Hirschwald, {J.\ Chem.\ Phys.}
\textbf{91}, 4949 (1989).

\bibitem{kris92} K.~Krischer, M.~Eiswirth, and G.~Ertl, {J.\ Chem.\ Phys.}
\textbf{96}, 9161 (1992).

\bibitem{ertl90} G.~Ertl, {Adv.\ Catal.} \textbf{37}, 213 (1990).

\bibitem{ojed12} C.~Ojeda and G.~M.~Buend\y a, {J.\ Comp.\ Meth.\ Sci.\ Eng.} {\bf 12}, 261 (2012). 

\bibitem{BRUND84}
C.~R.\ Brundle, J.~Behm, and J.~A.\ Barker, J.\ Vac.\ Sci.\ Technol.\ A {\bf 2}, 1038 (1984).

\bibitem{JAME99}
E.~W.\ James, C.~Song, and J.~W.\ Evans, J.\ Chem.\ Phys.\ {\bf 111}, 6579 (1999).

\bibitem{LIU99}
D.-J.\ Liu and J.~W.\ Evans, Phys.\ Rev.\ Lett.\ {\bf 84}, 955 (2000).

\bibitem{wint96}
J.~Wintterlin, R.~Schuster, and G.~Ertl, {Phys.\ Rev.\ Lett.} \textbf{77}, 123 (1996).

\bibitem{alba94} E.~V.~Albano and V.~D.~Pereira, {J.\ Phys.\ A: Math.\ Gen.} \textbf{27}, 7763 (1994).

\bibitem{khan04} K.~M.\ Khan and K.~Iqbal,  {Surf.\ Rev.\ Lett.} \textbf{11}, 117 (2004). 

\bibitem{mats79} T.~Matsushima, H.~Hashimoto, and I.~Toyoshima, {J.\ Catal.} \textbf{58}, 303 (1979). 

\bibitem{mach05a} E.~Machado, G.~M.~Buend\y a, P.~A.~Rikvold, and R.~M.~Ziff, {Phys.\ Rev.\ E}
\textbf{71}, 016120 (2005).

\bibitem{tome93} T.~Tome and R.~Dickman, {Phys.\ Rev.\ E}
\textbf{47}, 948 (1993).

\bibitem{buen06} G.~M.~Buend\y a, E.~Machado, and P.~A.~Rikvold, {THEOCHEM} \textbf{769}, 189 (2006).

\bibitem{seo08} H.~O.~Seo, S.~H.~Jeong, H.~J.~Lee, H.-G.~Lee, J.-H.\ Boo, D.~C.~Lim, and Y.~D.~Kim,  
{App.\ Catal.\ A}  \textbf{347}, 112 (2008).

\bibitem{bust00} V.~Bustos, R.~O.~U\~nac, and G.~Zgrablich, {Phys.\ Rev.\ E} \textbf{62}, 8768 (2000); 
{J.\ Mol.\ Cat.\ A} \textbf{167}, 121 (2001).

\bibitem{schm01} D.~H.~Schmidt and M.~Santos, {Phys.\ Stat.\ Sol.\ (a)}  \textbf{187}, 305 (2001).


\bibitem{hua03} D.Y.~Hua, F.~Zhang, and Y.~Q.~Ma,  {Phys.\ Rev.\ E}, \textbf{67}, 056107 (2003).


\end{thebibliography}
\end{document}